\documentclass[reprint,twocolumn,floatfix,preprintnumbers,prb ,nofootinbib,hyperref,include graphics]{revtex4-2}
\usepackage[T2A,T1]{fontenc}
\usepackage{amsmath}
\usepackage{amssymb}
\usepackage[utf8]{inputenc}
\usepackage{graphicx}
\usepackage{makecell}
\usepackage{diagbox}
\usepackage{amssymb,amsmath}
\usepackage[dvipsnames]{xcolor}
\usepackage{float}
\usepackage[normalem]{ulem}
\usepackage{upgreek}

\usepackage[breaklinks=true,hidelinks]{hyperref}
\hypersetup{allcolors=[rgb]{0.0 0.0 0.6},linkcolor=[rgb]{0.75 0.05 0.05}}

\begin{document}
\title{\textbf{\normalsize{Practical photonic band gap structures for high frequency axion haloscopes}}}

\author{\normalsize{D. Goulart$^{1,\dagger,*}$}}
\author{A.M. Sindhwad$^{1,\dagger,*}$}
\author{H.M. Jackson$^{1,\dagger}$}
\author{N.I. Kowitt$^1$}
\author{K.A. Dones$^1$}
\author{P. Castaño Basurto$^1$}
\author{A. Dawes$^1$}
\author{S. Jois$^1$}
\author{S.M. Lewis$^2$}
\author{K. van Bibber$^1$}

\affiliation{$^1$Department of Nuclear Engineering, University of California, Berkeley, Berkeley, CA 94720 USA}
\affiliation{$^2$Department of Physics and Astronomy, Wellesley College, Wellesley, MA 02481 USA}

\date{\today}

\begin{abstract}
Current and future searches for dark matter axions, based on their resonant conversion to photons in a magnetic field, span many orders of magnitude. A major impediment to designing resonators at the high end of this range, 5 GHz and above, is the proliferation of TE modes, which overwhelm and hybridize with the TM$_{010}$ mode to which the axion couples, making the search impossible. We demonstrate that a photonic band gap structure can be designed that completely suppresses the TE spectrum, even reducing the number of lattice periods to two or one, and violating perfect lattice symmetry. This allows tunable resonators to be designed in a convenient, volumetrically efficient circular geometry thus enabling future searches in the post-inflation axion mass range. 
\end{abstract}

\maketitle

\footnotetext[1]{dtgoulart@ucdavis.edu}
\footnotetext[1]{aaravsindhwad@berkeley.edu}
\footnotetext[2]{These authors contributed equally to this report.}

\section{Introduction}
The axion, a hypothetical ultralight particle, represents both the best solution to the Strong-CP problem \cite{PecceiQuinnPRL1977,WeinbergPRL1978,WilczekPRL78} and an ideal dark matter candidate \cite{Preskill1983,abbott1983cosmological, DINE1983137}. The search for the axion is daunting however, due to its extraordinarily weak couplings and the extraordinarily broad mass (frequency) range which must be scanned. Slow-roll inflation permits axions of mass as low as a 10$^{-9}$ eV \cite{graham2018stochastic,takahashi2018qcd}. On the other hand, in the cosmological scenario where the Peccei-Quinn (PQ) symmetry breaking responsible for the axion occurs after inflation, the mass is very high but is exactly calculable if it is assumed that axions saturate the dark matter density of the universe, $\Omega_{\text{DM}}$ $\sim$ 0.27. A recent calculation favors an axion mass within 45 - 65 $\upmu$eV, while allowing it to be as high as 300 $\upmu$eV \cite{6v21-d6sj}; another recent calculation calculation predicts a significantly higher bound, 95-450 $\upmu$eV \cite{Saikawa_2024}.

\begin{figure}[htbp]
    \centering
    \includegraphics[width = 1\columnwidth]{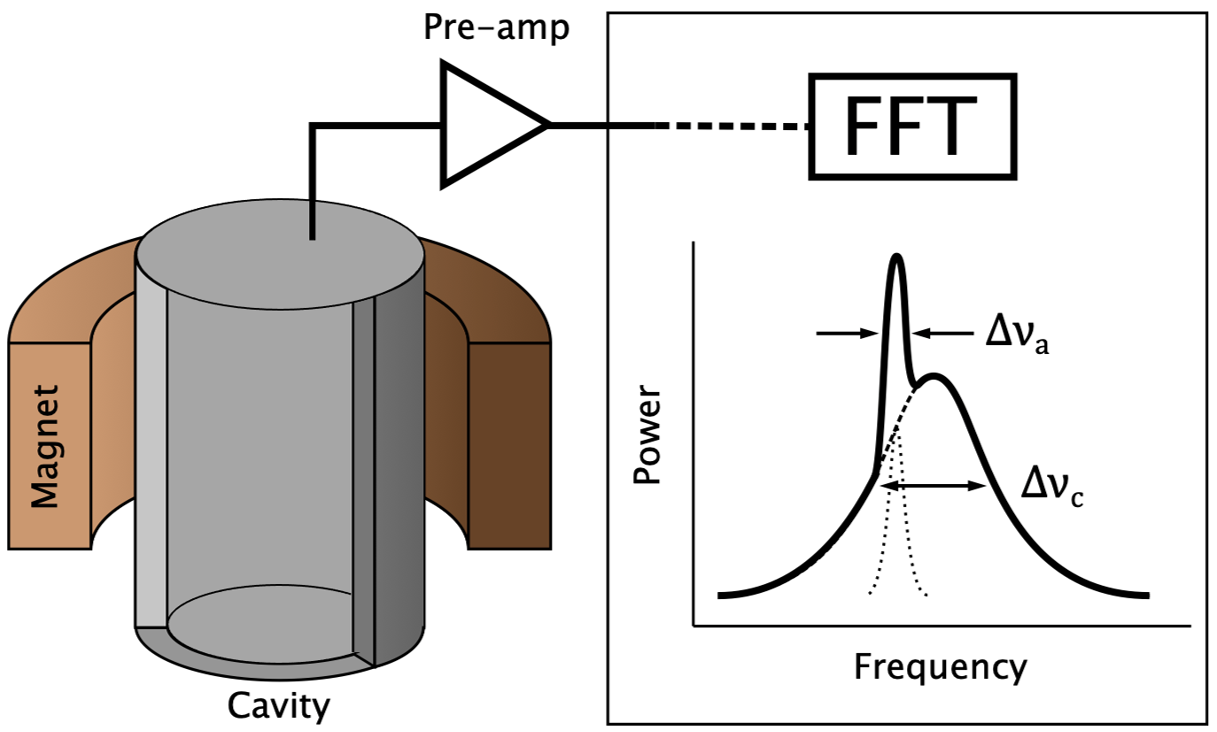}
    \caption{Schematic of the microwave cavity dark matter axion search (adapted from ~\cite{kowitt2023tunable}), showing a narrow axion signal on top of the broad cavity bandpass.}
    \label{fig:Fig 1 Microwave Cavity Schematic}
\end{figure}

The most sensitive method to search for dark matter axions is the resonant conversion of an axion to a photon in a tunable microwave cavity in a magnetic field (Figure \ref{fig:Fig 1 Microwave Cavity Schematic}). The resonant conversion condition is that the frequency of the mode is equal to the rest mass of the axion plus its small kinetic energy, $h\nu$ $=$ $mc^2(1 + O((v/c)^2))$ where $v$/$c$ $\approx$ 10$^{-3}$. In a traditional haloscope, the axion couples most strongly to the TM$_{010}$ or TM$_{010}$-like mode, but does not couple to TE or TEM modes. Since the original conception of the microwave-cavity axion experiment \cite{sikivie1983experimental}, most cavity searches have centered around the 2 $-$ 20 $\upmu$eV range, with a few outliers at higher masses \cite{AxionLimits}. Reducing the cavity size to reach higher frequencies incurs a strong penalty in cavity volume and thus conversion power, compromising the ability to detect the very weak signals expected, \textit{O}(10$^{-24}$ W). As a result, most experiments with sensitivity to the QCD axion have thus employed cavities of dimension greater than 100 mm, and consequently were limited to frequencies $<$ 6 GHz ($\approx$ 25 $\upmu$eV).

Several concepts have been explored for higher frequency cavities that do not sacrifice volume. Phase-combining multiple tunable cavities has been demonstrated to work (up to four) but represents a complex engineering challenge \cite{S.D.Kinion.Thesis}. Cavities incorporating a tunable symmetric lattice of metal rods have been developed, and such a cavity is currently being prepared for operation in HAYSTAC \cite{Simanovskaia_2021}. These can extend the frequency range from that of more conventional cavities by an octave or so, but a new paradigm is needed to increase the range by an order of magnitude or more.
Recently, a promising concept to reach much higher frequencies has been put forward, the plasma axion haloscope based on wire-array metamaterials \cite{lawson2019tunable}; research demonstrating how its plasma frequency can be engineered and tuned has been encouraging \cite{millar2023searching,Wooten2023,kowitt2023tunable}.


\begin{figure}[htbp]
    \centering
    \includegraphics[width = 0.8\textwidth, angle = 270]{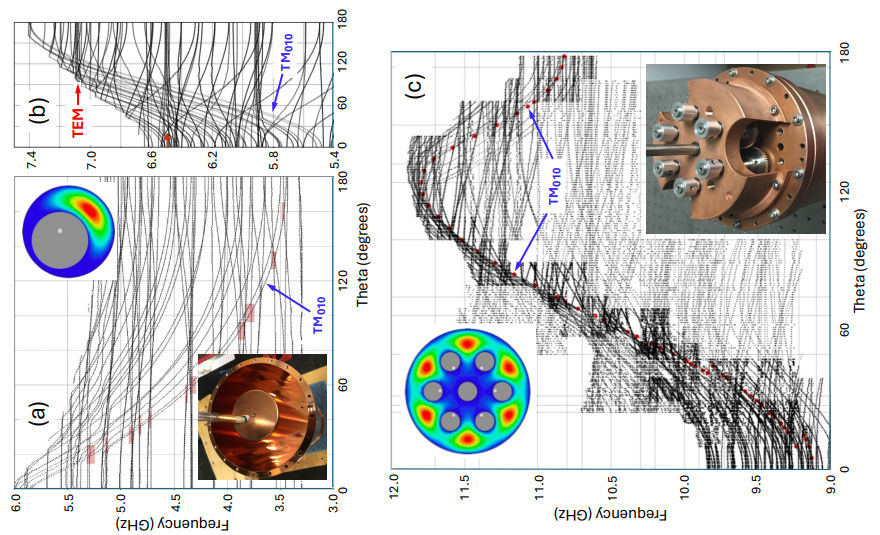}
    \caption{Simulations  of eigenmode frequency vs. tuning rod angle for three cavity configurations, all of length $L$ $=$ 254 mm. (a) Cavity of radius $R$ $=$ 50.8 mm, with a single tuning rod of radius $r$ $=$ 25.4 mm. \cite{al2017design} (b)  Cavity of radius $R$ $=$ 50.8 mm, with seven tuning rods of radius $r$ $=$ 7.9 mm \cite{Simanovskaia_2021}.  (c)  Same as (b) but with $R$ $=$ 39.6 mm.  The TEM manifolds are most clearly seen in (b), uniformly spaced every $c/2L$ $=$ 590 MHz.}
    \label{fig:Fig 2 mode maps}
\end{figure}

Increasing the frequency of a haloscope must not incur an undue penalty scanning speed, which in terms of the main experimental parameters is proportional to $B^4V^2C^2Q$, where $B$ is the magnetic field, $C$ is the form factor for the mode of interest (effectively a measure of the overlap of the cavity mode and the external magnetic field), $V$ is the volume of the cavity and $Q$ is its quality factor \cite{Peng:2000hd}. Cavities for axion dark matter searches should also have a useful dynamic range in frequency to be practical, ideally 10\% or more. There is one additional criterion for a cavity to be functional, namely that the mode of interest, usually the TM$_{010}$ mode for a cylindrical cavity, should be relatively free of mode-mixing, or hybridization with the TE and TEM modes it encounters as it is tuned across its range. While in principle crossing modes need not mix, in physical cavities any perturbations whether unintended (e.g. resulting from finite dimensional and alignment tolerances) or designed (e.g. antenna ports, tuning axles) will generally lead to mixing at some level. The degree of hybridization on the analysis may be tolerable, leading to no loss of frequency coverage, to severe, where several MHz of data will be unusable in the crossing region. In the latter case a secondary run must be performed with some change of configuration to move the crossing far enough away such that the frequency interval can be rescanned. The greatest challenge in the search for the post-inflation axion is the rapid proliferation of TE modes with increasing frequency. Figure \ref{fig:Fig 2 mode maps} represents the simulated mode maps for three HAYSTAC cavities, spanning from 3.4 GHz to 11.6 GHz, with the frequency loss going from tolerable to so severe that the TM$_{010}$ mode cannot be identified.

As a result of mode-crossings, some frequency intervals must be cut from the data analysis. These intervals can be determined by the field profiles of the modes as they approach one another, which directly manifests the degree of hybridization \cite{al2017design,Rapidis_2019,Simanovskaia_2021}.

Being able to identify detrimental mode crossings however, is insufficient in itself; what is required is the ability to mitigate mode mixing to the greatest degree possible. Photonic band gap (PBG) structures represent a promising solution to the problem. A PBG is a regular lattice, triangular or square, of metal or dielectric rods of radius $a$ and spacing $b$, for which there are both propagating solutions and band gaps in the photon dispersion relation. Further, by removing one or more rods in the middle of such an array, one can form a void, and a judicious choice of $a$ and $b$ can trap TM modes in the void while radiating away TE modes. PBG structures have made a significant impact in optical photonics, and have also been studied for their potential utility in microwave sources and linear accelerators \cite{Kroll:1992xu,smirnova2002simulation,smirnova2005fabrication}. We have simulated and measured PBGs to explore their potential to eliminate intruder TE modes in a microwave cavity axion experiment. 

A PBG resonator with a single metallic tuning rod was first constructed. Its performance was identical to simulations \cite{lewis2024tunablephotonicbandgap}; the TM$_{010}$ mode was cleanly identified across its entire frequency range, 7.4 $-$ 9.5 GHz, as the spectrum was cleansed of TE modes. The PBG does not eliminate TEM modes, which are spaced every $\Delta \nu$ $=$ c/2L, 590 MHz for all cavities discussed here. 

\begin{figure}[htbp]
    \centering
    \includegraphics[width = 1\columnwidth]{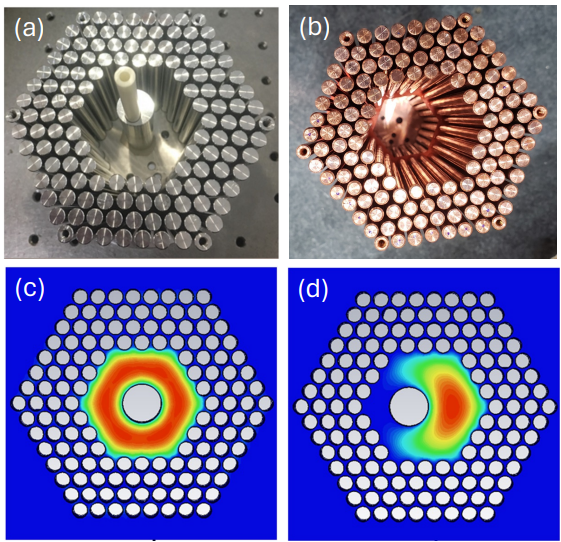}
    \caption{PBG structures comprising 132 rods ($a$ = 3.2 mm), with 37-rod void \cite{lewis2024tunablephotonicbandgap} (a) Aluminum (94 mm length), with tuning rod.  (b)  Copper (254 mm length), without tuning rod.  (c), (d) Simulations of the TM$_{010}$ E-field showing full containment of the mode in the removal void, with the tuning rod on- and off-axis respectively.}
    \label{fig:Fig3 hexpbg}
\end{figure}

That resonator, however, would be unsatisfactory as an axion haloscope, as the void confining the TM modes represents less than a quarter of the total available volume (Figure \ref{fig:Fig3 hexpbg}).

The purpose of the present work is to further explore the practicality of incorporating PBGs into higher frequency axion haloscopes. To this end, a resonator similar to that of Ref.~\cite{lewis2024tunablephotonicbandgap}
was built incorporating several improvements from the previous study. (i)  First, as the key operating parameter of the PBG is the ratio of rod radius to spacing, $a$/$b$, the radius and spacing of the rods were proportionately reduced to increase the sensitive volume of the resonator. (ii)  The PBG was designed as four concentric circles with the same number of rods, to further increase the interior volume of the resonator. (iii) To study how thin the PBG could be made, and thus increase conversion volume, a study was done in which successive rows of the PBG were removed, ultimately leaving a single row. (iv)  Finally, the tuning scheme consisted of a symmetric lattice of six movable rods, which represents the first step towards a metamaterial-based tunable haloscope in the future.  

\section{Description of design and construction}

We wish to emphasize the entirely pragmatic purpose of this study: to explore how far one can deviate from an ideal photonic band gap structure to maximize resonator volume yet still suppress the TE spectrum.  While the validity of a band theory description of finite thickness structures of exact symmetry is well established \cite{smirnova2005novelThesis}, it is clearly inappropriate for curved structures of only one or two rows.  For simplicity, however we will retain the terminology photonic band gap structure (PBG) throughout, as it morphed from it.

This study was carried out with the resonator described in \cite{Simanovskaia_2021}, with the exception that the barrel representing a closed boundary condition was replaced with a circular PBG structure, representing an open geometry. Tuning was accomplished by using a hexagonal array of 6 movable metal rods of 7.9 mm radius (the fixed central rod was removed for this study), extending the full length of the cavity, 254 mm, excepting gaps of $\sim$150 $\upmu$m between the ends of the rods and the top and bottom end caps.  The rods are made of aluminum electroplated with copper and annealed. The rods move radially by rotating around an off-axis pivot driven by a common gear mechanism maintaining hexagonal symmetry at all pivot angles.


\begin{figure*}[htbp]
    \centering
    \includegraphics[width=1\textwidth]{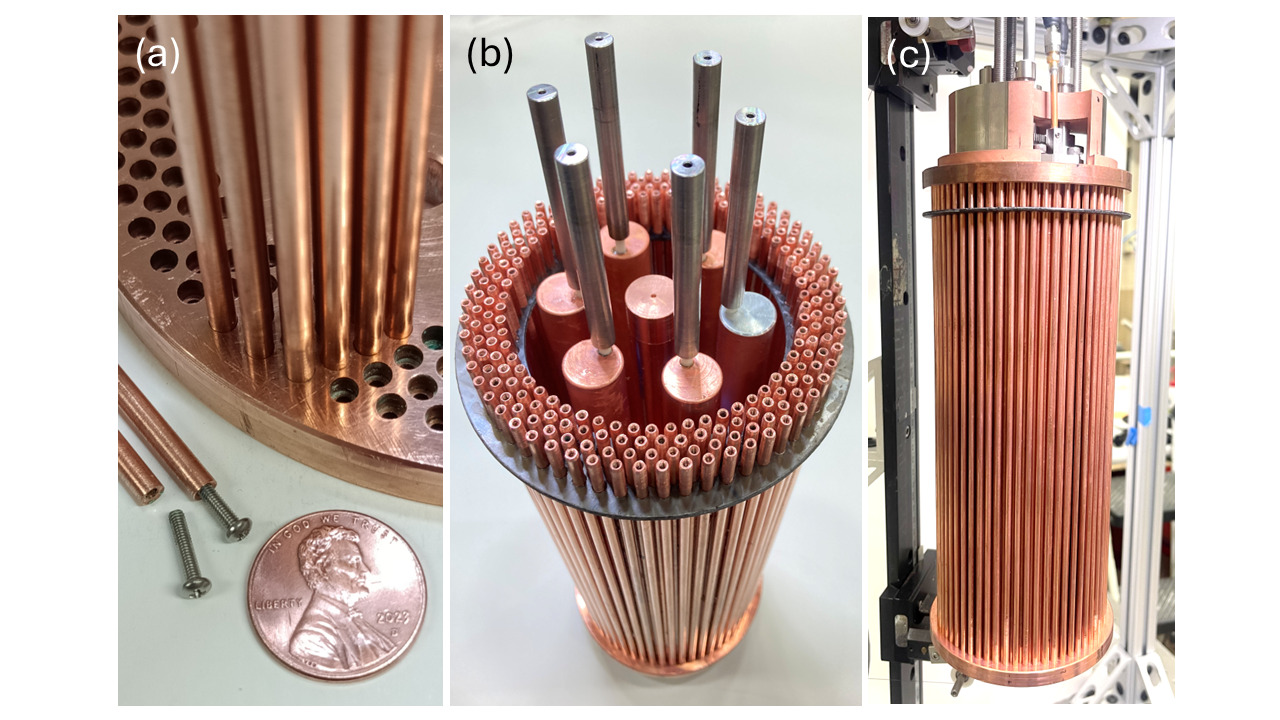}
    \caption{(a)  Detail of how the rods constituting the PBG structure are affixed to the end caps.  (b)  The resonator with the four-row PBG and the hexagonal array of tuning rods in place, with the top end cap and gearbox removed.  (c)  Fully assembled resonator mounted on test stand. The photo here shows a tuner of 6 movable rods with one fixed rod in the center; all the data in this paper were taken in the configuration of 6 movable rods, with no central rod.}
    \label{fig:Fig 4 7rodPBGAssembly}
\end{figure*}

The PBG structure (Figure \ref{fig:Fig 4 7rodPBGAssembly}) consisted of 4 circular rings of 60 OFHC copper rods of 3.18 mm diameter, the radii of the four rings being 41.7, 44.8, 48.0 and 51.2 mm.  The 240 rods were beveled on each end to leave a small circumferential knife-edge to maximize metal-to metal contact upon compression, and thus minimize resistive losses for microwave surface currents between the PBG rods and end caps.  The rods were also drilled and tapped on each end to accommodate 1-72 screws which provided the compression of the thin rods within their recesses in the OFHC copper end caps.  To facilitate the final assembly of the resonator, a 3D-printed thin HDPE alignment grid was used to ensure each rod would find its socket, which was left in place during the tests.  Furthermore, it was also found necessary to sand down a slight conical shape to one end of the rods, which had the effect of introducing a non-uniformity in length of the rods of up to 100 $\upmu$m.

While rings of an equal number of rods do not constitute a perfect triangular lattice (here $a$/$b$ ranging from 0.36 to 0.30 from the inner to the outer ring), if the spacing of the rods relative to the radius of the ring is small, the array may still effectively confine TM modes while suppressing TE modes. Confirming this behavior was a key research goal of this work. 

\section{Empty resonator studies}

\begin{figure*}[htbp]
    \centering
    \includegraphics[width = 1\textwidth]{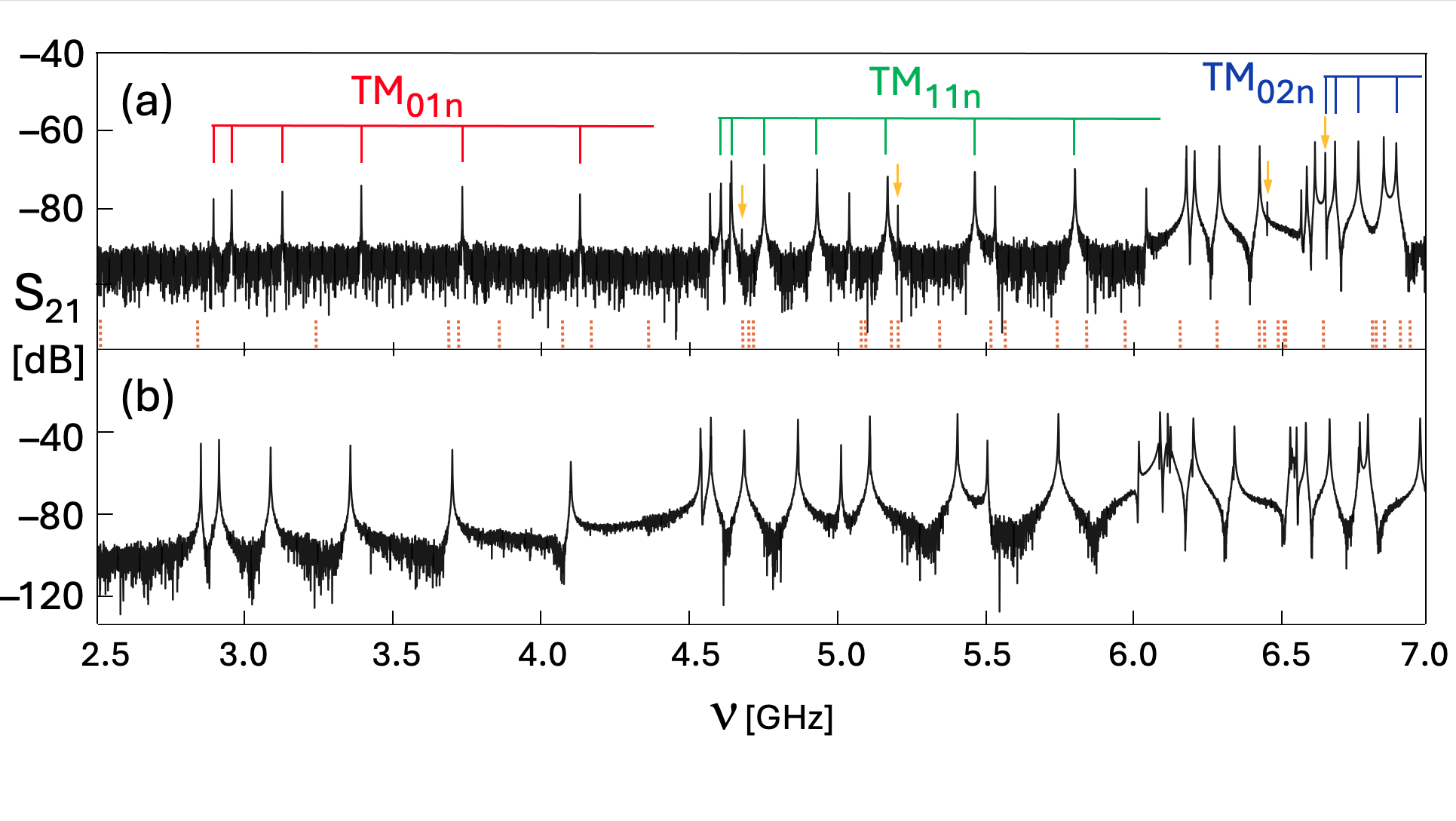}
    \caption{(a)  Spectrum where the 4-row PBG has been replaced by a copper cylinder, thus constituting a conventional microwave cavity. (b) The spectrum of the 4-row PBG structure, with no tuning rods inside.  The corresponding peaks are slightly shifted downwards in frequency as the inside diameter of the cylinder was slightly smaller than the effective diameter of the PBG, as described in the text.  In (a), the orange dotted lines indicate the calculated frequencies of all TE modes within that frequency range; the yellow arrows indicate TE modes that were visible with the cylindrical boundary condition but were eliminated with the PBG.}
    \label{fig:Fig 5 BarrelvPBGspectrumcomparison}
\end{figure*}

The resonator was first tested without any tuning rods in place, to evaluate the effectiveness of the PBG in confining the TM$_{010}$ mode by itself. Figure \ref{fig:Fig 5 BarrelvPBGspectrumcomparison} shows the spectrum of the PBG along with the spectrum where the PBG has been replaced by a cylinder, the latter thus constituting a conventional cylindrical cavity.  The inner radius of the cylinder was beveled to a knife edge for maximum metal-to-metal contact with 20 threaded rods providing the compression between the end caps and the barrel. The radius of 39.7 mm was chosen to be slightly less than that of the inner surface of the PBG rods (40.1 mm) to ensure full contact with the end cap, without the knife-edge contact being interrupted by the sockets of the inner row of the PBG.  This resulted in the spectrum with the cylinder being shifted slightly higher in frequency than that of the PBG.

The study of the empty PBG resonator exhibited several important systematics.  First, the measured frequency of the TM$_{010}$ mode was identical to that of a cavity with a cylindrical boundary condition (i.e. a continuous conducting tube), as analytically calculable \cite{jackson1962}, for an effective radius $R_{\text{eff}}$ $=$ $R - r$, where $R$ $=$ the radius of the resonator measured to the center of the first row of PBG rods and $r$ being the radius of the PBG rods itself, here being $R_{\text{eff}}$ $=$ 41.7 $-$ 1.6 mm $=$ 40.1 mm.  Furthermore, the measured frequencies of all TM modes up to 7 GHz agreed with the frequencies calculated with $R_{\text{eff}}$ within a fractional error of $\sim$10$^{-3}$. In other words, the PBG resonator behaved identically to that of a cylindrical cavity whose radius is circumscribed by the innermost surface of the stockade of rods.  The same pertained to the TE modes where they were visible. TE modes exist across the entire spectrum but generally are not strongly excited unless they are close enough in frequency to a TM mode to mix with it, the antenna being oriented perpendicular to the electric field, and mounted just within the end cap where the E-field vanishes (see for example Figures 10 and 11 in Ref. \cite{Simanovskaia_2021}.

Second, the frequencies of the TM modes did not depend on the number of rows, N of the PBG, even for N=1.  N=1 occupies the innermost ring, with N$>$1 building out from there.  


\begin{figure}[htbp]
    \centering
    \includegraphics[width = 1\columnwidth]{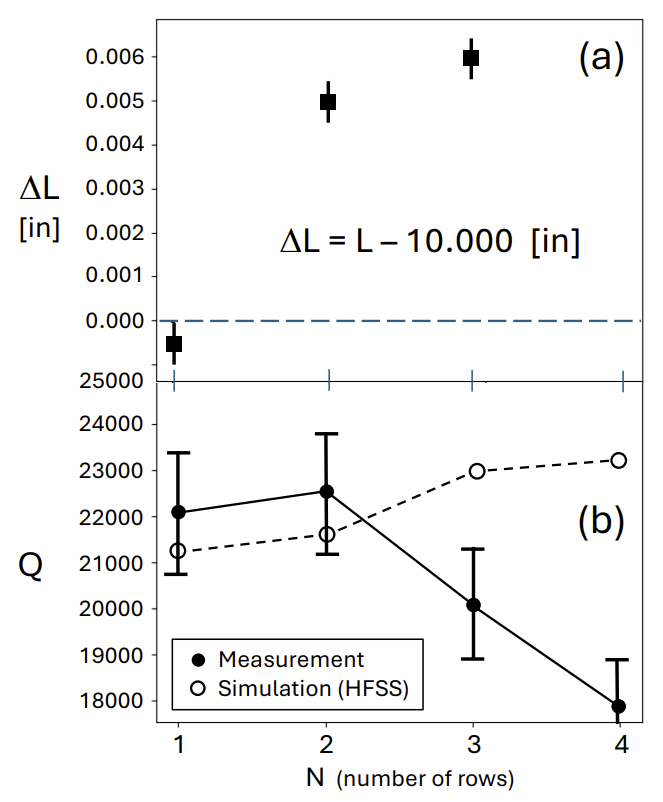}
    \caption{(a) Distance measured between the end caps of the resonator.  The design separation set by the lengths of the PBG rods in their sockets is 10.000 inches.  The measured distance for $N$ = 4 was lost but was not less than that for $N$ = 3.  (b) Measured and simulated $Q$ for the empty PBG structure, as a function of the number of rows in the PBG.  }
    \label{fig:Fig 6 QcomparisonEmptyCavity}
\end{figure}

Third, the unloaded quality factor $Q$ for the TM$_{010}$ mode was measured at room temperature and compared with the quality factor simulated with HFSS \cite{Ansys}, as a function of $N$ (Figure \ref{fig:Fig 6 QcomparisonEmptyCavity}, bottom).  For the PBG with even just one row of rods with $a$/$b$ = 0.364, the measured $Q$ of 22,100 was consistent with the calculated value within measurement error. That the TM$_{010}$ mode should be effectively confined and the $Q$ limited only by surface resistance was confirmed by performing the simulation in the perfect electrically conducting (PEC) limit, where the $Q$ would reflect only radiative loss. In fact, the radiative quality factor for N = 3 is Q$_{rad}$ = 4.5 $\times$ 10$^8$ and even for N = 1, Q$_{rad}$ = 8 $\times$ 10$^7$.  While the PEC simulations were performed only to demonstrate that radiative loss from the TM$_{010}$ mode through the stockade of rods is completely subdominant to the resistive losses for OFHC copper, it may be helpful in understanding such a surprisingly high radiative quality factor to consider a closely related geometry for which there exists an analytical solution, i.e. parallel-polarized transmission through a planar array of rods in the limit where the diameter of the rods is comparable to their spacing \cite{article, 1140011}.  For our specific parameters, $r$ = 1.59 mm, $a$ = 4.37 mm, $\nu$ = 2.87 GHz ($\lambda$ = 104.5 mm), the calculated transmission through a single row is 6.2 $\times$ 10$^7$, favorably comparing with 8 x 10$^7$ for the HFSS simulation of the curved array.

As mentioned, the effective diameter of the PBG, as determined by fitting its spectrum, is equal to that of a cylinder circumscribed by the innermost surface of the first row of the PBG.  However, the quality factor for a copper cavity with that effective diameter would be 28,400, and one may thus wonder about the discrepancy between the cavity's value and that of the PBG of $\sim$23,000, given that radiative losses are entirely negligible.  We address this question by recalling that the quality factor for a cylindrical cavity can be written as

\begin{equation}
    Q = \left(\mathrm{Geometrical \,\, Factor}\right) \left( \frac{V}{S\cdot\updelta} \right),
\end{equation}

where $V$ is the volume of the cavity, $S$ is the total cavity surface area, and $\updelta$ is the skin depth. For all TM modes, the geometrical factor is identically 2 \cite{jackson1962}.  While the effective diameter of the resonator may be that of the innermost surface of the first row, the PBG rods represent a corrugated surface whose effective area is larger than that of a circular cylinder. Assuming the effective surface area is defined by hemicylinders facing the interior, the quality factor can be estimated to be 24,400, which is in reasonable agreement with both the HFSS calculation and experimental measurements.

As subsequent rows are populated, the measured $Q$ is observed to rise slightly for $N$ = 2, but then falls off for $N$ = 3, 4.  This can be ascribed to a mechanical issue, namely that in assembling the PBG with a larger number of rods, it became increasingly difficult to fully compress the resonator, as evidenced by differences in the distance between the end caps measured at several points around the circle, as shown in Fig.~\ref{fig:Fig 6 QcomparisonEmptyCavity}.  As the conical knife edges lose contact with the bottom of their respective wells, the joint resistance goes up, thus degrading $Q$.  Ideally one would braze the rods into their end caps, but the requirement of being able to reconfigure the tuning rods multiple times drove us to a design where the resonator was screwed together.  An optimal mechanical design for flexible reconfiguration without sacrificing the quality factor remains a challenge for future work.

\section{Studies of a Multirod - PBG Resonator}

The tunable multirod cavity was chosen for these measurements as it represents a first step toward larger lattice and metamaterial-type resonators, relevant to searches for the post-inflation axion \cite{millar2023searching}. Here we have taken the cavity in Ref. \cite{Simanovskaia_2021} and replaced the cylindrical tube with a cylindrical PBG to investigate its efficacy in suppressing TE modes at higher frequencies.  

For these measurements, the PBG was configured with the inner two rows populated; the resonator itself consisted of six movable tuning rods, with gaps between the ends of the tuning rods and the end caps of approximately 150 $\upmu$m. For comparison, data was also taken with a smaller OFHC copper tube (radius 39.7 mm), which approximates the effective radius of the PBG.

As before with an empty resonator, we have now determined the mathematically equivalent radius of the PBG with a multirod tuner at a much higher frequency, by varying the radius of a cylinder in simulation to match the measured frequency of a TM$_{010}$ mode of a single-row PBG at 8.925 GHz.  We again conclude that the electromagnetic mode terminates sharply at the radius of the inner surface of the PBG, the difference here being 0.03 mm, at the level of machining tolerances for the resonator. Thus, the volume and form factor can be calculated exactly as for a cylindrical cavity of the same radius.

\begin{figure*}[htbp]
    \centering
    \includegraphics[width = 1\textwidth]{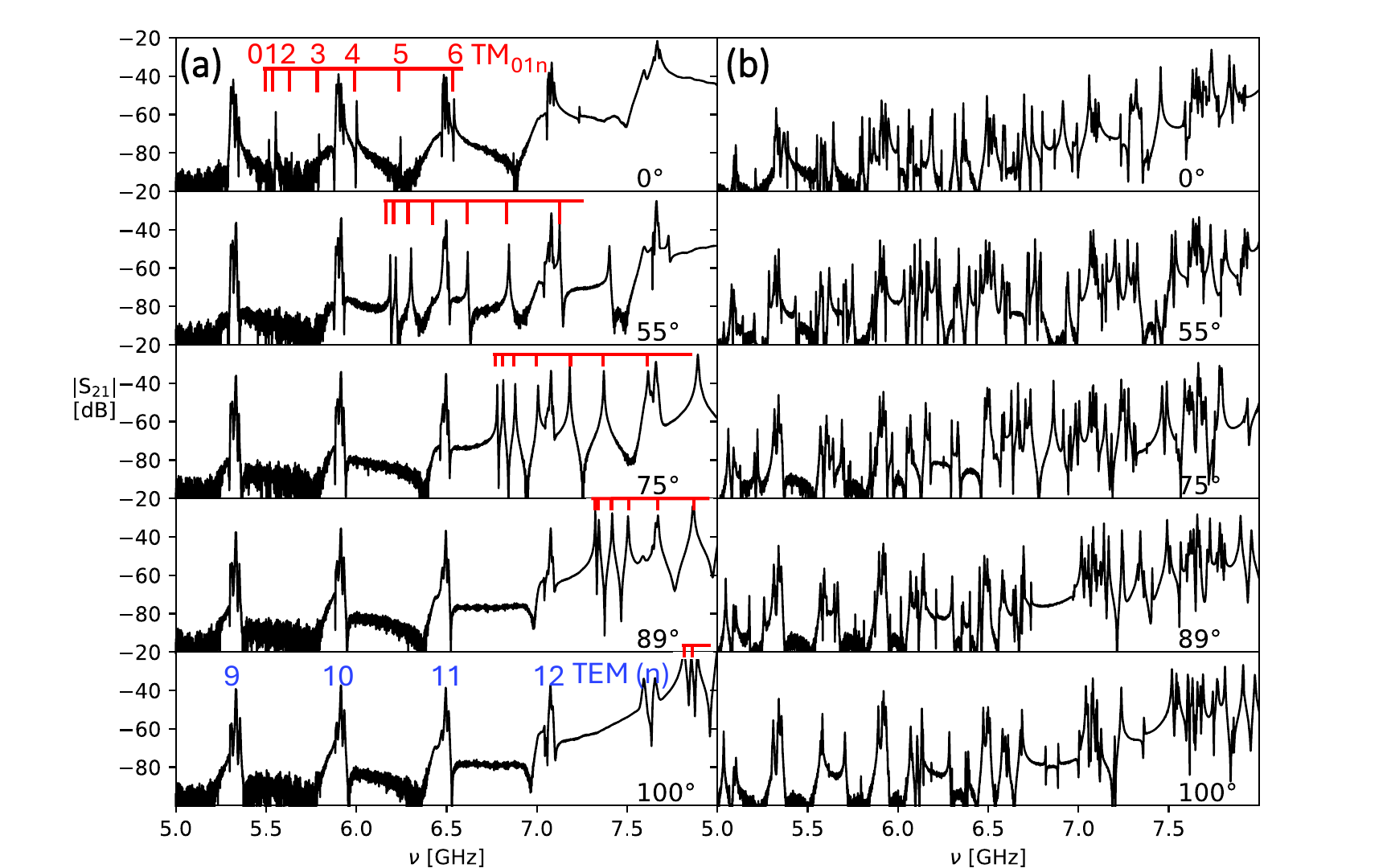}
    \caption{(a) Spectra of the multirod resonator with the PBG at various pivot angles of the tuning rods; $\theta$ = 0$^{\circ}$ corresponds to the hexagonal array of rods in the maximally outboard position.  (b)  The corresponding spectra where the PBG has been replaced with a cylindrical copper tube.}
    \label{fig:Fig 7}
\end{figure*} 

Figure \ref{fig:Fig 7} shows selected spectra at various pivot angles of the tuning rods both for the cylindrical boundary condition and the PBG structure. For the PBG, the TE modes have been completely eliminated, leaving only the TEM modes and the low-order TM modes. The TM$_{010}$ mode of interest and the co-moving TM modes are clearly resolved up to 8 GHz, where they are gradually lost among higher-order TM modes which move downward as the rods move outward (the design of the resonator was only optimized for the lower range of frequencies).  The frequency scan was automated at a fixed cavity coupling which led to the TM modes being under-coupled at lower frequencies and over-coupled at higher frequencies. As a consequence, only at the lowest frequencies where the resonator is most weakly coupled could the quality factors of the resonator with the PBG and with the cylinder be meaningfully compared, at 5.552 GHz being 11,500 and 10,420 respectively. The corresponding spectra with the cylindrical tube are also shown in Figure \ref{fig:Fig 7}; the spectral density is so great that the TM$_{010}$ mode is completely unresolved, precluding its use as an axion haloscope.

\begin{figure}[htbp]
    \centering
    \includegraphics[width = \columnwidth]{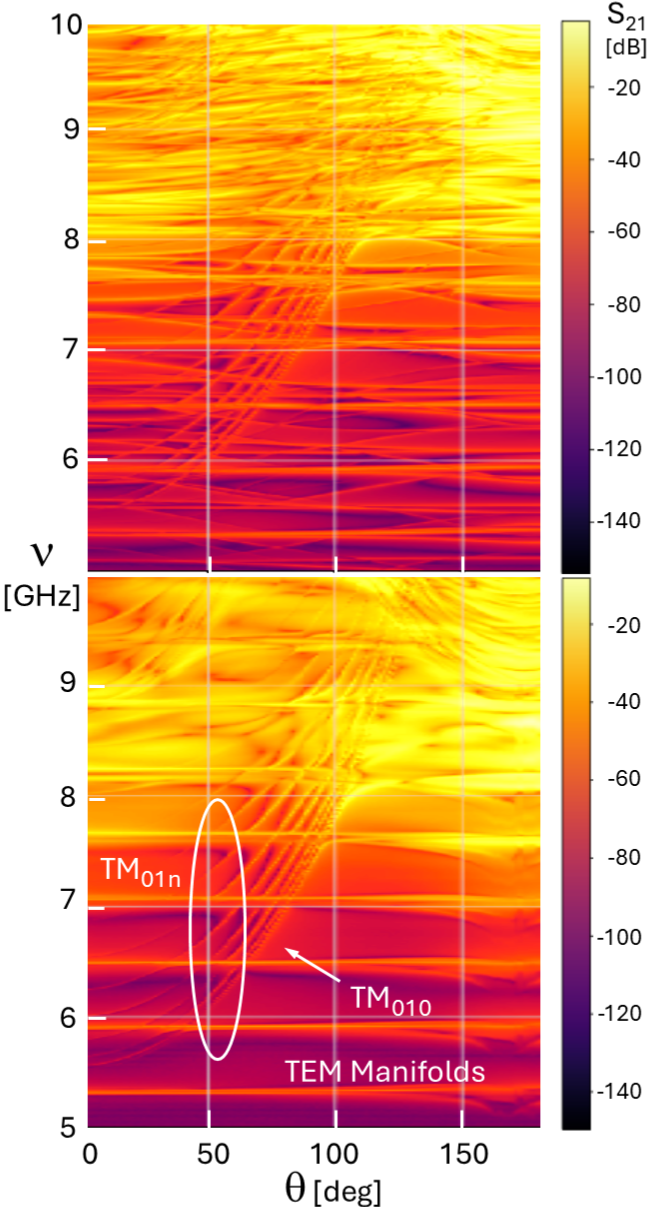}
    \caption{Mode maps of the six-movable rod symmetric resonator.  (Top) Conventional microwave cavity, where a copper cylinder provides the radial boundary condition, E$_z$(R) $=$ 0.  (Bottom) PBG structure representing a pass band for TE and stop band for TM modes.}
    \label{fig:Fig 8}
\end{figure} 

The same data is represented by the mode map in Figure \ref{fig:Fig 8}, which dramatically indicates how the PBG structure suppresses the spectra of TE modes. 

As the PBG does not eliminate the Transverse Electric and Magnetic (TEM) modes, it may be wondered why the TEM modes are broad, and what could be done about them.   For multiple internal tuning rods there are as many TEM modes as there are tuning rods, 6 in the present case.  In simulations, the TEM modes in the manifold are exactly degenerate at the frequencies $\nu$ $=$ $\frac{nc}{2L}$ when the tuning rods are in contact with the top and bottom end caps.  For finite gaps, the modes split.  The splitting increases linearly with the width of the gaps, and furthermore increases as the rods move closer together, as reflected in the TEM manifolds occupying $\sim$10\% of the spectrum at smaller angles and $\sim$20\% at larger angles.
 
The width of the TEM can thus be minimized by reducing the gaps, and it should be remembered that all modes will narrow at cryogenic temperatures as the $Q$ improves, typically by a factor of 3-4 from room temperature for a copper cavity. A straightforward way of recouping all the frequency lost to
TEM modes is simply to perform a follow-on run with a shorter or longer resonator that would move the TEM modes away from the regions to be scanned.

\begin{figure}[htbp]
    \centering
    \includegraphics[width = 0.85\columnwidth]{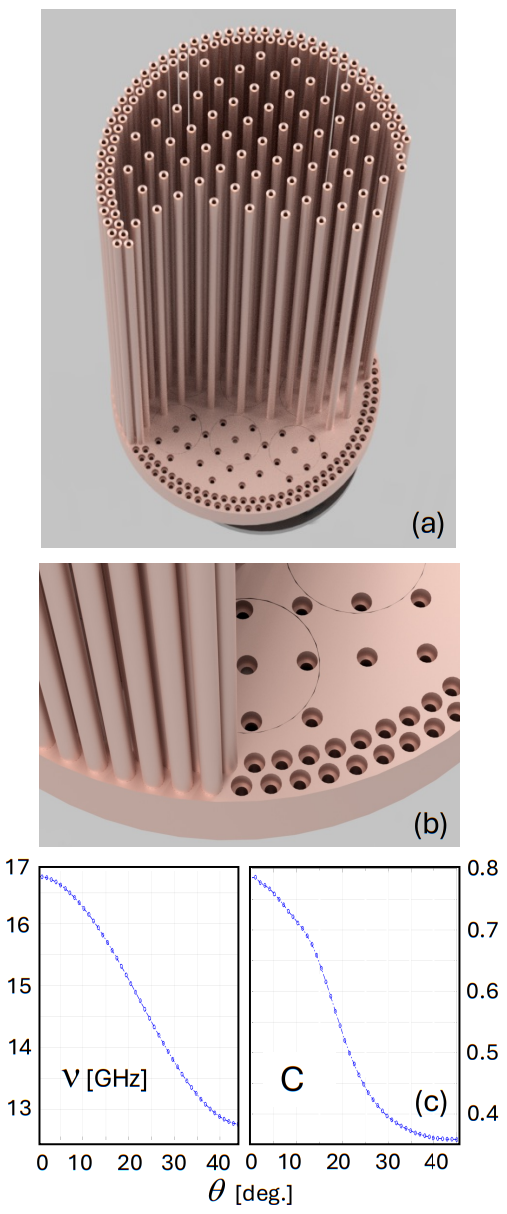}
    \caption{(a) Design of a prototype resonator for the post-inflation axion search experiment, ALPHA, under construction \cite{millar2023searching}.  (b) The detail focuses on a two-row stockade of rods for TE mode suppression, with the same $a$, $b$ as the innermost two rows here.  The prototype is designed to HAYSTAC format, i.e. 10.2 cm I.D. and 25.4 cm height; the full ALPHA resonator will be $\times$1.75 larger in linear dimensions. (c) The frequency $\nu$  and form factor $C$ for the prototype resonator. The frequency and tuning of the resonator are determined by 44 fixed rods and 9 pods of 5 rods each which rotate in concert, the angle of rotation designated as $\theta$.}
    \label{fig:Fig 9}
\end{figure} 

This test resonator was not intended as an engineering prototype, but only to determine the minimum number of rows that would contain the TM$_{010}$ mode of interest while suppressing TE modes. As one or at most two rows has proven adequate, the PBG structure for an actual axion haloscope would be located at the maximum possible radius determined by the bore of the magnet, and need not take up any more radial space than a conventional cylindrical barrel. Figure \ref{fig:Fig 9} represents a prototype resonator under construction for the ALPHA post-inflation axion search.

An early operational deployment of a multi-rod resonator with a PBG boundary will be carried out in the HAYSTAC experiment in the near future \cite{PhysRevLett.134.151006}.

\section{Summary and future work}

This work represents a first step toward practical resonators at higher frequencies motivated by the search for the post-inflation axion. A lattice-type tuning scheme has been coupled with a photonic band gap structure and has demonstrated that the TE forest can be eliminated, thus largely preventing mode hybridization and sharply diminishing spectral coverage at higher frequencies.  Excellent results have been obtained with structures of only 2 rows or even 1 row of rods, further improving the volume utilization of a resonator within the given geometrical constraints. 

An important conclusion of this work is that the electromagnetic mode terminates sharply at the radius of the inner surface of the PBG. Thus, the volume and form factor can be calculated exactly as for a cylindrical cavity of the same radius.

Some practical implementation issues remain to be worked out. The current method of assembling the PBG resonator involves screwing the rods into their respective holes. With this method it is more difficult to fully compress the resonator for a larger number of rods.  A design in which the rods could be permanently brazed into the end caps while preserving the ability to assemble and disassemble the resonator would be optimal, and options are being examined.  Furthermore, in the experiment, the resonator would reside within the innermost thermal shield separating the resonator from the superconducting magnet. It may be wondered whether the thermal shield, a metal tube only slightly larger in diameter than the resonator would represent a secondary cavity which would support TE modes and reflect power back into the resonator, thus defeating the purpose of the PBG.  A first simple test was performed roughly simulating an actual haloscope.  A long thermal shield of approximately the diameter of that of HAYSTAC and closed at the bottom was raised up over the resonator suspended from its test stand.  No change whatsoever was observed in the spectrum, but it will be important to repeat this exercise in greater fidelity to the experiment.  The third issue again pertains to the thermal shield.  The resonator will be suspended from the mixing stage of the dilution refrigerator, which cools it to a base temperature $T$ $<$ 100 mK.  The innermost thermal shield, however, will be at a temperature of $T$ $\sim$ 700 mK, thus defining the blackbody environment within the shield.  Given the open geometry of the PBG, it remains to be seen if the effective noise temperature, as measured in the receiver chain reflects any anomalous thermal component indicative of contamination from the blackbody photon spectrum. Wrapping microwave absorbing material around the resonator may prove helpful with the latter two issues.

\section{Acknowledgements} 
This work was supported by the National Science Foundation, Grant No. PHY-2209556. 

\clearpage

\bibliography{bibliography}

\end{document}